# The Decline of Online Knowledge Communities: Obstacles, Workarounds, and Sustainability

CHING CHRISTIE PANG, The Hong Kong University of Science and Technology, China
XUETONG WANG, The Hong Kong University of Science and Technology, China
YUK HANG TSUI, Hong Kong University of Science and Technology, China
PAN HUI, The Hong Kong University of Science and Technology (Guangzhou), China

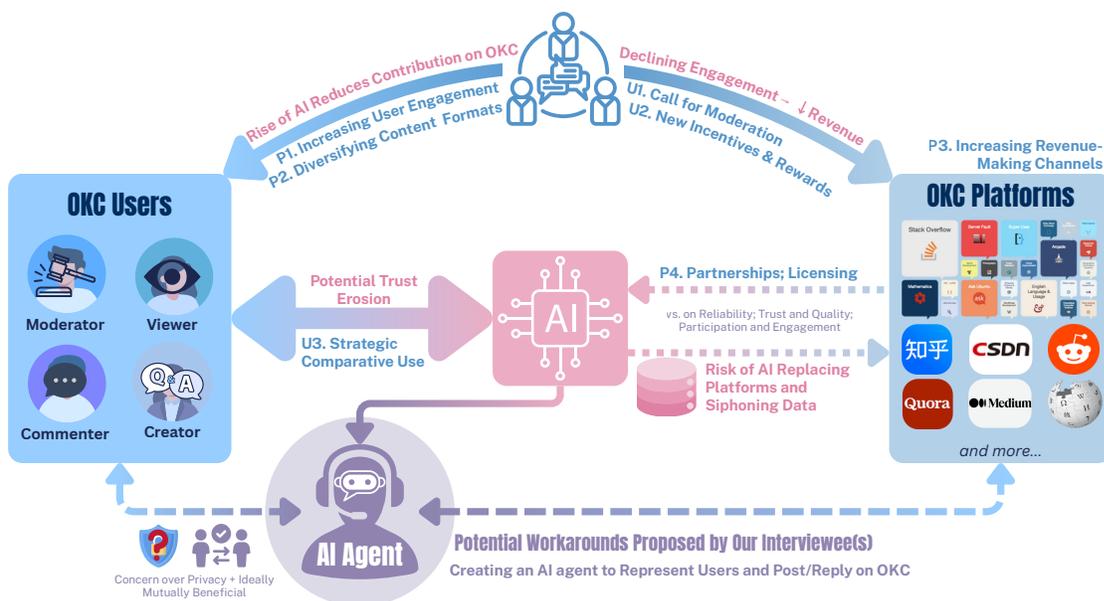

Fig. 1. Obstacles and Workarounds of the Rise of AI and Impact (*potential decline*) on Online Knowledge Communities (OKC) during Knowledge Seeking in Complex, Real-World Settings. Text in pink refers to obstacles, blue refers to workarounds, and purple refers to our proposed workarounds. U1-U3 depict users' workarounds, and P1-P4 refer to platforms' workarounds.

Online knowledge communities (OKC) such as Stack Exchange, Reddit, and Zhihu have long functioned as socio-technical infrastructures for collective problem solving. The rapid adoption of Generative AI (GenAI) introduces both complementarity and substitution.

Authors' Contact Information: Ching Christie Pang, The Hong Kong University of Science and Technology, Hong Kong SAR, China, ccpangaa@connect.ust.hk; Xuetong WANG, The Hong Kong University of Science and Technology, Hong Kong SAR, China, xwangdd@connect.ust.hk; Yuk Hang Tsui, Hong Kong University of Science and Technology, Hong Kong SAR, China, yhtsui@connect.ust.hk; Pan Hui, The Hong Kong University of Science and Technology (Guangzhou), Guangzhou, China, panhui@ust.hk.

Large language models (LLMs) offer faster, more accessible drafts, yet divert traffic and contributions away from OKC that also provided their training data. To understand how communities adapt under this systemic shock, we report a mixed-methods study combining an online survey (N=217) and interviews with 11 current users. Findings show that while users increasingly rely on AI for convenience, they still turn to OKC for complex, ambiguous, or trust-sensitive questions. Participants express polarized attitudes toward AI, reflecting divergent hopes and uncertainties about its role. Yet across perspectives, sustaining sociability, empathy, and reciprocity emerges as essential for community resilience. We argue that GenAI's impact constitutes not a terminal decline but a design challenge: to reimagine socio-technical complementarities that balance automation's efficiency with human judgment, trust, and collective stewardship in the evolving knowledge commons. To decline or sustain, it is now or never to take action.

## 1 Introduction

Online knowledge communities (OKCs) are socio-technical systems where people ask questions, contribute answers, and curate resources at scale [20], evolved by various social question-and-answer (Q&A) platforms [71, 76]. Exemplars include Stack Exchange[1], Quora[2], Reddit subcommunities [3], and Zhihu[4]. Kuang et al. define these sites as places where individuals concurrently share knowledge, experience, and expertise through questions and answers [40], scaffolding learning, encode community norms, and sustain collective problem solving [29, 54]. For HCI, they are both public infrastructures for reliable knowledge and living testbeds for studying incentives, moderation, reputation, trust, and socio technical governance [53].

Generative artificial intelligence (GenAI) has introduced a systemic shock to these communities. Large language models (LLMs) such as ChatGPT now perform information seeking and content generation with striking proficiency [10, 69]. Effects are heterogeneous. On the complementarity side, LLMs provide immediate drafts and broaden access, with evidence of increased question novelty by casual users after ChatGPT's release [61]. On the substitution side, LLMs attract knowledge seekers away from communities, depressing visits, posting, and answer diversity [10, 21, 69]. Complicating matters, OKCs seeded early training data for today's models [1]. [35] discuss how the misinformation from AI affect users on Stack Overflow. Declines in fresh community content risk constraining future training corpora, creating a feedback loop that weakens both the communities and the models that depend on them. We provide relevant examples and case studies in Section 2.1 for a more comprehensive understanding of these issues.

This situation reflects the infrastructural issues identified in human-computer interaction (HCI) by Edwards et al., where technical design limits user outcomes [22]. AI's automation enhances efficiency but restricts participatory and collaborative opportunities [5]. Design choices prioritizing immediacy hinder joint inquiry and interpretation, creating a mismatch for the slow deliberation in online knowledge exchanging process. Despite evidence of declining participation and shifts in novelty [10, 61, 69], guidance for sustaining communities amidst AI integration is lacking. While this has raised discussion in HCI, existing studies focus on the discussion over a more concentrated roles [42] and platforms

---

[1] https://stackexchange.com/  [2] https://www.quora.com/  [3] https://www.reddit.com/  [4] https://www.zhihu.com/

Manuscript submitted to ACM

[36, 43]. As such, there is a gap in actionable insights on user practices and platform strategies to mitigate substitution, preserve community value, and foster a symbiotic relationship between OKCs and AI.

This paper addresses that gap by comparing user experience across AI and OKCs during everyday knowledge seeking (section 4.1). We identify obstacles (section 4.2), analyze user and platform workarounds (section 4.3), so as to explore different roles' perceptions of these issues (section 4.4). Our mixed-methods study includes an online survey of 217 participants and interviews with 11 experienced users. Findings reveal a shift towards AI for convenience, though with lower confidence in its accuracy compared to OKCs. While participants express confidence in recognizing AI, it potentially increase cognitive load, and therefor we call for attention on AI literacy. Participants anticipate continued volume decline in OKCs due to AI's rise, expressing trust in human-generated knowledge while acknowledging community limitations. Users and platforms adapt workflows to address challenges, advocating for stronger moderation and specialized niches. Interviewees suggest a hybrid model where AI assists privately and shares vetted contributions back to the community. Most users developed their own strategies, while platforms are exploring new business models to stabilize revenue amidst these changes. However, the platforms' workarounds may unintentionally lead to potential decline due to their violation of the communities' reciprocity. We also find that different roles in OKCs indicate different levels of engagement; respondents also held different perceptions on the impact of AI in OKCs. We speculate that respnodents with higher particpation levels are more inclined to adopt stricter regulatory and labeling measures, or even reject the user of AI in any online platforms; while those with lower engagement levels support a survival-of-the-fittest strategy.

This work makes two main contributions. **First**, it empirically maps how knowledge seekers reallocate tasks between AI convenience and community trust, revealing a situated identifier on obstacles; synthesizing user and platform workarounds, highlighting design levers for reliability, provenance, and incentive alignment. **Second**, we also articulate a design space that treats AI and OKCs as interdependent infrastructures rather than substitutes, outlining governance and product directions to stabilize the knowledge commons. Above all, the paper reframes decline as a design problem of sustaining socio-technical complementarities necessary for public, reliable knowledge production.

## 2 Background and Related Work

We first provide background on the decline of OKCs (section 2.1), supported by empirical evidence and examples. We then review related work on the motivations and dynamics of OKCs (section 2.2 and 2.3), which is crucial for understanding their operation and evolution. This exploration shows that AI's impact on OKCs (section 2.4). Our study is motivated by these complexities within the HCI, CSCW, and social computing literature.

### 2.1 Background: the Declining Communities

The decline of OKCs is evident through traffic losses and structural disruptions. For example, Stack Overflow saw new questions plummet from 108,563 in November 2022 to 25,566 in December 2024, a 76.5% drop coinciding with the launch of ChatGPT [5]. This trend aligns with a reported 50% loss in site traffic [6] following ChatGPT's release though this data was regarded as inaccurate according to the official announcement [7]. Nevertheless, even before to the AI trend, Stack Exchange Inc. had already laid off 28% of its workforce in October 2023 due to reduced user engagement [8], suggesting that AI acts as an accelerant rather than the sole cause of this decline.

---

[5] GitHub Gist Stack Overflow Statistics by *hopeseekr*   [6] Medium by *ThreadSafe Diaries* and Linkedin Post by *Gregor Ojstersek*   [7] Blog by Stack Overflow on Traffic (August 9, 2023)   [8] A Message from Prashanth Chandrasekar, CEO Stack Overflow (October 16, 2023)

The introduction of new AI-generated content (AIGC) policies across various platforms has sparked intense discussions among users, with some reporting that their answers were incorrectly deleted [9]. In response, Reddit's r/AskHistorians and r/AskScience subreddits even banned AI submissions. Furthermore, ambiguity surrounding AI policies led to the 2023 Moderation Strike [10], while a partnership with OpenAI prompted protests over data usage in 2024 [4]. Concerns regarding Reddit's API pricing and user data mining have further intensified perceptions of data injustice and declining community trust. Similar declines, such as the closure of Yahoo Answers in 2021, indicate a broader shift extends far beyond AI as a single factor. Understanding how OKCs adapt amid automation is thus crucial for designing resilient, human centered knowledge platforms.

### 2.2 Motivations for Knowledge Sharing Online

Knowledge sharing is central to Online Knowledge Communities (OKCs), supporting the collective creation and exchange of expertise. Users share when perceived benefits such as reputation, reciprocity, or belonging outweigh costs like time and effort [52, 55, 71, 76]. Gamification effectively enhances such motivation [14, 17].

Social Exchange Theory (SET) explains these behaviors by viewing participation as an exchange driven by expected rewards and reciprocity [6, 46]. In OKCs, intangible payoffs—reputation and social recognition—promote active and timely contributions [73]. Because shared knowledge is a public good, continued engagement relies on fairness, reputational growth, and mutual obligation rather than material incentives [7].

From a social-cognitive view, individuals imitate rewarded peers; vicarious learning and herd behavior lead new users to follow high-reputation members [2, 47]. Complementary theories highlight utilitarian benefit, enjoyment, and self-efficacy as sharing motivators [18, 41]. Building on this, Preece's empathic community model [59] and Wenger's communities of practice (CoP) [77, 78] emphasize learning through shared endeavors and emotional reciprocity.

Empirical findings confirm that reciprocity expectations, reputation systems, and peer feedback sustain participation [13, 51, 63]. Intrinsic motivation remains strongest when supported by visible community rewards [85]. Understanding these motivational patterns is crucial, as AI now transforms knowledge sharing by reshaping engagement norms and redefining trust within OKCs.

### 2.3 From Motivating to Interaction: the Dynamics of Online Knowledge Communities

Transitioning from motivations to the structure and dynamics of OKCs, it is crucial to examine how these motivations appear in community interactions. OKCs such as Reddit, Stack Overflow, and Zhihu differ in moderation systems and technical affordances that shape organization and knowledge exchange [34, 75]. Interface and algorithmic designs influence discussion flow: voting mechanisms, or collaborative filtering [15], rank content by popularity, which may reinforce dominant views and overlook minority perspectives [49]. Threaded discussions on Reddit and Zhihu foster continuity through nested replies [9], while Stack Overflow's structured Q&A format values accuracy and conciseness but limits open-ended dialogue.

Online communities support collective sense-making by helping users reframe problems and build shared understanding [83, 84]. OKCs form part of broader online ecosystems [67], with cross-posting and shared membership across platforms [74]. Most questions receive quick replies [58], though conversations often favor common knowledge over rare expertise [70], and information accuracy evolves through user interaction [19].

---

[9] Example from Stack Exchange: posted by *Starship* (August 2024)  [10] Original Post on Moderation Strike by *Tinkeringbell* (June 2023)

Moderation—"governance mechanisms that structure participation to enable cooperation and prevent abuse" [26]—is essential for sustaining community health [54]. Prior studies introduce blocklists [24], expression filters [32], and narrative moderation [82]. Oliveira et al. [54] found that removal explanations, whether from humans or bots, help users grasp community norms and reduce future post removals. Because moderation rules vary across platforms, self-governance, volunteered, and elected moderation remain common in platforms like Reddit and Stack Overflow.

Volunteer moderators and experienced users act as gatekeepers of community standards. Based on participation typologies, Xu and Ye [80] identified browsing, initiating, and participating as key sharing modes. Members may be posters, contributors, or lurkers [76], driven by self-efficacy and resource accessibility [30]. Other frameworks distinguish seekers and sloths [23], or contributors and customers [65]. Kou et al. also categorized the social roles into knowledge broker, translator, conversation facilitator, experienced practitioner, and learner [38]. Guided by these models, we choose the umbrella terms to explain to our respondents and categorize roles as **moderators**, content contributors (**creators**/**commenters**), and content users/**viewers**. These roles collectively structure knowledge flow in OKCs, a process increasingly reshaped by AI-driven participation and evolving notions of expertise and trust.

### 2.4 The Influence of GenAI on Online Knowledge Communities

GenAI and AI, particularly large language models (LLMs) such as ChatGPT, is reshaping how participants create and validate knowledge within OKCs. This technological shift introduces both innovation and disruption—enhancing efficiency and accessibility while challenging authenticity, collaboration, and trust. Recent studies highlight that AI's growing presence alters user behavior, content quality, and community engagement, calling for new governance mechanisms to sustain participation [10, 25, 31]. Studies also examine how Reddit communities and moderators adapt AI-related rules, highlighting the rise of subreddit policies on generative content [42] and documenting moderators' challenges in maintaining quality and autonomy amid limited detection tools [43].

Empirical evidence demonstrates both benefits and strains. The landmark study by Burtch et al. analyzed Stack Overflow and Reddit developer communities, finding a 12% decline in daily visits and over a 10% drop in posted questions after ChatGPT's release [10]. Socially oriented forums such as Reddit's discussion subcommunities appear more resilient due to stronger interpersonal ties and belonging [10]. Although LLMs can improve individual productivity, they also generate hallucinated or misleading content, undermining mentorship and collective learning [61]. Within organizational OKCs, similar patterns could weaken attachment, innovation, and career development. Sustaining engagement therefore requires embedding AI as an assistive yet not substitutive tool and reinforcing moderation to preserve information quality [33, 64].

Parallel research connects AI to new challenges of trust and authenticity. Controlled experiments reveal AI's capability to deceive humans [57], raising concerns of misuse by malicious actors [25, 48]. Studies in AI-mediated communication [28] further show that merely suspecting AI involvement reduces interpersonal trust [31, 62]. At the same time, generative tools complicate existing issues such as harassment and synthetic media abuse[16, 56, 72]. While existing studies focus on the problem as a phenomenon [10, 25] or concentrate on specific roles within communities [42, 43], we aim to explore the broader implications of GenAI on the dynamics of trust and participation across entire OKCs. These developments highlight an urgent need for transparent AI disclosure, adaptive moderation, and clear authorship norms. Addressing this gap is crucial for understanding how OKCs can maintain trust, expertise, and active participation as GenAI becomes integral to their knowledge ecosystems.

## 3 Methods

Inspired by the growing influence of GenAI on OKCs under different stakeholders, we explored the following research questions:

- **RQ1**: How do users compare their experiences with AI and OKCs, and what are the impacts of AI on OKCs (i.e., obstacles and workarounds)?
- **RQ2**: How do different roles within OKCs perceive the influence of AI?
- **RQ3**: What are the design implications for enhancing OKCs in the context of AI?

### 3.1  Data Collection: Survey (RQ1 and RQ2)

To gather a diverse range of insights, we adopted methods informed by Baumer et al. [3] and Lu et al. [44], who used online survey-based methodology to collect stories about issues related to Facebook non-use and live streams in China respectively. These studies demonstrated that such methodologies can effectively gather social platform data to analyze larger trends and events. Given that one of our study's contributions is to discuss the entire landscape of OKCs, which includes platforms primarily in English and Chinese, we developed an online questionnaire available in two languages, validated by proficient English and native Chinese authors, to ensure clarity and cultural relevance in the choice of words.

*3.1.1 Survey Design and Review of Platforms.* The survey comprised 50 questions, organized into three distinct thematic sets, followed by a demographic section which collect information including participants' careers, regions, and invitations for follow-up interviews.

*i. Experience with AI Tools and Its Impact on OKC Participation.* Here, we first focused on participants' experiences with AI tools. It assessed how often they use these tools, their levels of trust, and their perceptions of content creation and originality (Questions 1–6). We also looked at the impact of AI tools on participation in social Q&A and OKC platforms. It compared user behaviors before and after the adoption of AI and identified changes in motivation and participation (Questions 14–30). Additionally, it measured attitudes toward AIGC. Participants rated their trust, perceived quality, bias, ethics, and effects on motivation using a 7-point Likert scale (Questions 31–42).

*ii. Roles and Practice in OKC.* This set of questions examined respondents' roles and practices within OKCs. It gathered information about their experience, engagement roles, and the platforms they actively use (Questions 7–13).

*iii. Strategies and Workarounds for Sustaining OKCs.* This section investigated strategies for sustaining OKCs in the age of AI. It assessed perceived risks and preferred interventions, such as moderation, integration, or incentives (Questions 43–48). We also confirmed our review on platforms' workarounds and open-ended questions to collect the firsthand stories and experiences from participants.

The survey utilized a mix of question types, including multiple-choice, Likert-scale, and open-ended items. It also included two attention-check questions (Question 21 and 37), ensuring data validity and participant reliability. Ethical considerations were paramount; participants were informed about the study's purpose, and consent was obtained prior to participation. Anonymity and confidentiality were assured throughout the data collection process.

In addition to the survey, we conducted a platform review to analyze how platforms have responded to the use of AI. We gathered data from public announcements, platform policies, and user discussions on platforms including *Stack Overflow* and *Stack Exchange*, *Reddit*, *Quora*, and *Zhihu*. Key incidents were highlighted and their introduction is illustrated in section 2.1. Our analysis involved systematic searches of platform archives and verified media sources. The collected materials were thematically coded to identify patterns in governance, community sentiment, and adaptation strategies related to AI content integration in OKCs, later to be confirmed with interviewees to formulate the platforms' workarounds.

*3.1.2 Recruitment and Screening.* To recruit a diverse participant pool and avoid our results being biased by any single platform of users, we conducted regular postings on various OKC, including Zhihu and subcommunities such as *r/SampleSize*, *r/SurveyExchange*, and *r/AskReddit* on Reddit, over a six-week period. We also recruited through *r/ArtificialInterlligence* after requesting to the moderator. Besides, we promoted the survey on commercialized platforms like *surveyswap.io* and posted flayers within the authors' institution, employing purposive and convenience sampling to target more OKC users.

We followed the data handling methods outlined by Haimson et al. [27]. Initially, we received 264 responses. After manual inspection, we removed 47 responses due to failure in answering the attention check questions (N=23), insufficient completion time (N=11), and patterned or nonsensical responses (N=13). This left us with 217 completed responses for analysis.

*3.1.3 Profile of Respondents.* The questionnaire required approximately 15 to 20 minutes to complete (M = 16.20 minutes, SD = 10.34) . We intentionally did not collect gender or other personally identifiable demographic data to minimize irrelevance and protect privacy. Respondents reported a wide range of professional identities, including students, researchers, software developers/engineers, practitioners, public-sector employees, professionals in applied domains, and educators. The final sample consisted of respondents from various regions, including Asia (N=123), Europe (N=53), North America (N=26), South America (N=6), Africa (N=5),  and four participants who chose not to disclose their location.

Figure 2 visually summarizes the distribution of respondents' profiles. Respondents were categorized based on their participation in knowledge-sharing activities, aligning with prior classifications of seekers and sloths [23], and contributors and customers [65]. Most respondents were from Asia and were mainly motivated by information seeking and knowledge exchange, consistent with prior findings from China and India [81].

## 3.2 Data Collection: Semi-Structured Interviews

*3.2.1 Interview Recruitment.* To gather in-depth insights from a diverse range of users, we recruited 11 interviewees, including a moderator. Participants were selected through invitations embedded within the questionnaire. Respondents who expressed interest in follow-up interviews were contacted via email or WhatsApp for confirmation and asked to share their OKC profile pages to verify their actual engagement.

Recruiting active and willing OKC participants proved challenging, as many users were difficult to reach or hesitant to commit due to privacy concerns and time constraints. Some potential participants declined because they were no longer active in online communities or were uncomfortable discussing their use of AI. For instance, one moderator who initially expressed interest later withdrew, stating via email that he decided to leave the community and was unwilling to share his experiences. Despite these challenges, we successfully recruited 11 participants, encompassing a diverse range of OKC roles and regions.

*3.2.2 Participants.* Interviews were conducted in two rounds: the first in August 2025 and the second in November 2025. Five sessions were held face-to-face, while six were conducted remotely via Zoom or Google Meet, ensuring flexibility for participants in different locations. Each interview lasted approximately 20 to 70 minutes, and participants received an honorarium of HKD 50 via PayPal or FPS in appreciation of their time.

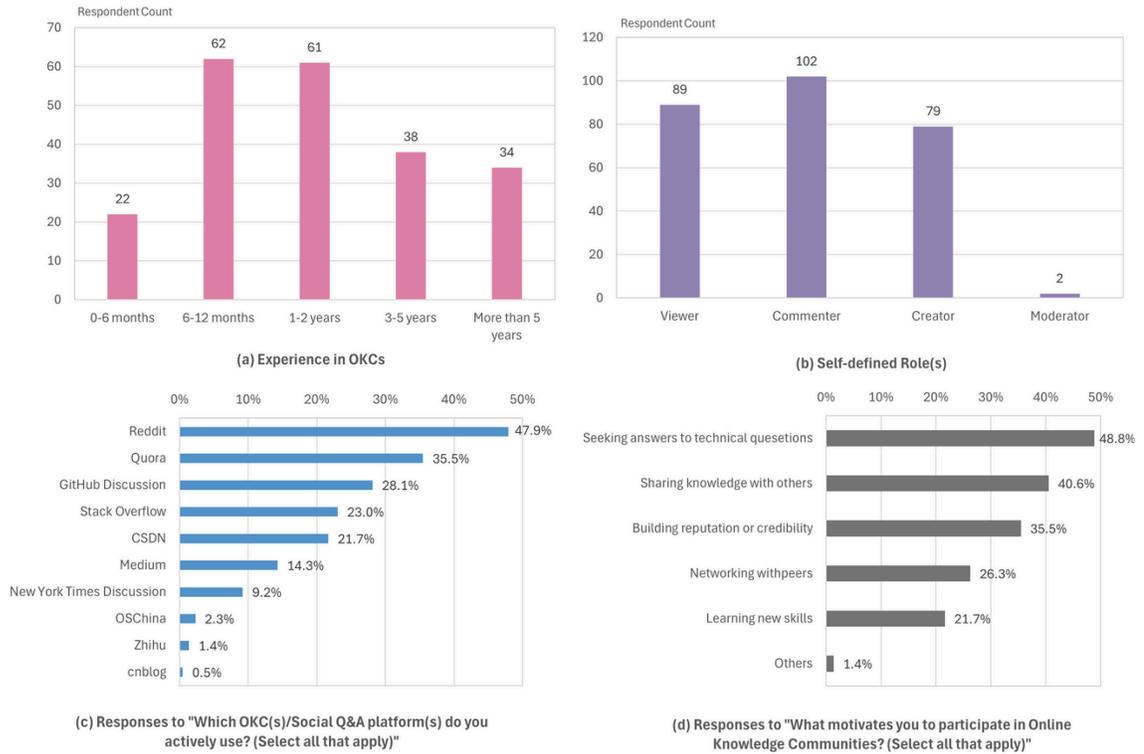

Fig. 2. Distribution of experience, self-defined roles, platforms actively used, and motivations with OKCs.

Semi-structured format was adopted to maintain consistency while allowing participants to elaborate freely. Different sets of questions were used depending on the participants' roles in their OKCs. Topics covered included current participation practices, perceived changes after AI integration, trust and attitudes in AIGC, and strategies or workarounds for adapting to evolving community dynamics.

Interviews were conducted in English, Mandarin, or Cantonese, depending on participants' language preferences. All sessions were audio-recorded, transcribed verbatim, and reviewed by two authors to ensure accurate representation of participants' voices and meanings. Table 1 summarizes the demographic and background of all interviewees.

### 3.3 Data Analysis

Due to non-normality [66], we used the Wilcoxon signed-rank test [79] for comparative analysis, as the data were ordinal. Additionally, we utilized factor analysis to group related questions, facilitating the identification of significant patterns within the data. Factor loadings were derived through principal component analysis and subsequently rotated using varimax rotation, following established methodologies [37].

| ID | Role | Years of Use | Region /Residence | Frequency of Use (times/week) | Most Commonly Used Platforms | AI Tools Usage |
|---|---|---|---|---|---|---|
| A1 | Viewer | 7 | Mainland China | 3–4 | Stack Overflow, Zhihu, CSDN | ChatGPT, Copilot |
| A2 | Viewer | 5 | Mainland China | 2 | Stack Overflow, CSDN | ChatGPT, DeepSeek |
| A3 | Creator | 10 | Hong Kong SAR | 7 | Quora, Stack Exchange, Stack Overflow, Zhihu, Reddit | ChatGPT, Claude |
| A4 | Viewer | 5.5 | Mainland China | 3 | Stack Exchange | DeepSeek |
| A5 | Creator | 6 | Hong Kong SAR | 6 | Stack Overflow, Quora | ChatGPT |
| A6 | Viewer | 4 | Singapore | 5 | Reddit, Stack Exchange | ChatGPT, Gemini |
| A7 | Commenter | 8 | India | 7 | Reddit | ChatGPT |
| A8 | Viewer | 9 | Germany | 1 | Stack Overflow, Quora, Reddit | None |
| A9 | Viewer | 5 | India | 2 | Reddit, Quora | ChatGPT |
| A10 | Moderator | 12 | United Kingdom | 5 | Stack Overflow, Reddit | Capilot |
| A11 | Creator | 11 | Canada | 7 | Reddit, Stack Exchange | None |

Table 1. Summary of interview participants' demographic and background information.

For the qualitative data, open-ended responses were analyzed using an open coding method. Two authors fluent in both Chinese and English independently coded the initial 20% of responses and then convened to reach a consensus on the coding scheme. One author subsequently coded the remaining responses, and both coders met again to ensure agreement on the codes. These codes were translated into English through manual translation and then verified with the assistance of LLM, followed by a review to compare and confirm appropriate terminology.

We collected about 9 hours of interview recordings, which were transcribed for analysis. Using a deductive thematic analysis [8], we based our coding on predefined themes while allowing new insights to emerge. Two independent coders first reviewed the transcripts line by line, developing initial codes that captured participants' responses. They then compared and discussed their codes, resolving differences and refining categories through several iterations. Finally, we used affinity diagramming to group related codes into thematic clusters [50], which helped identify broader patterns and insights about participants' experiences and perspectives on AI and OKCs.

### 3.4 Positionality Statement

The study was approved by the university's Institutional Review Board (IRB). The lead author has over eight years of active participation across multiple OKCs (i.e., Zhihu, Reddit, Quora, Stack Overflow, and Stack Exchange), holding various badges and memberships. All authors are experienced in AI tools. This long-term engagement informed the study design and interpretation while maintaining reflexive awareness of potential bias.

## 4 Findings

The findings are organized around the key themes that emerged from our mixed-method analysis combining survey and interview data. We first examine how users compare their experiences with OKCs and AI (**RQ1**), detailing the potential trade-off between efficiency and quality on each system. We then move from these comparisons to the broader challenges AI has introduced for OKCs (**RQ1**). Following these obstacles, we describe how both platforms and users have begun acknowledging and working around these emerging limitations (**RQ1 and RQ3**). Finally, we explore how different roles within OKCs perceive and respond to AI's influence (**RQ2 and RQ3**), illustrating how AI reshapes participation, credibility, and the sustainability of online knowledge ecosystems overall.

### 4.1 Comparing AI and OKCs (RQ1)

*4.1.1 User Participation and Engagement.* The integration of AI into everyday information-seeking has significantly altered user reliance on OKCs. Wilcoxon signed-rank test results confirmed substantial declines in key participation activities after the emergence of GenAI. Reported **comment-contributing** activity dropped largely (Before: M=3.37, Mdn = 3; After: M=2.48, Mdn = 2, Mdn = 3; $Z = -7.6$, $p < .001$, $r = -0.6$), along with **answer-contributing** (Before: M=3.25, Mdn = 3; After: M=2.99, Mdn = 3; $Z = -3.7$, $p < .001$, $r = -0.4$)) and **question-asking** (Before: M=3.78, Mdn = 4; M=3.18m, Mdn = 3; $Z = -5.8$, $p < .001$, $r = -0.5$), both showing medium-to-large effects. Collectively, these results show a broad reduction in participatory energy, echoing A10 who is also the moderator of Reddit, that *"AI makes finding things faster but makes communities quieter"* (A10). OKC remains valued but less frequented, indicating declining collective momentum and shifting attention patterns.

Patterns of engagement suggest a structural reordering of knowledge-seeking behavior . Many participants (A1-A7, A9) described AI as their "first resort" for fast or factual queries, reserving OKC for nuanced, community-specific, or experiential discussions. As A3 noted, *"I check ChatGPT for quick syntax, but for developers or bugs debates, I still read Stack Overflow threads."* A7 echoed that *"AI is great for simple translation or overviews, but when it's about design judgment, I go to people."* Many participants described a triangulation habit: beginning with AI for initial exploration, then verifying information through OKCs before applying it.

*4.1.2 Trust and Quality.* Trust remained a defining factor separating AI and OKC. Participants valued the lived experience, context awareness, and conversational feedback loops characteristic of OKC. AI's breadth and clarity were acknowledged but seen as insufficient without transparency, provenance, and accuracy. A7 summarized this tension that *"I use AI first when I need a rough map, but I ask people when I want to know if it actually works... AI answered random stuff as it only wants to echo and agree with me."* Similarly, A5 observed that *"AI's answers sound confident... but without examples, I don't believe them. OKC posts come with reputations, votes, and back-and-forth."* A2 also recalled, "AI answers are fast, but I can't verify who said it. On Stack Overflow, at least I can read the discussion and comments below." Note that A8 and A11, who had never used AI before, exhibited complete distrust and opposition toward AI due to concerns about its quality and hallucination.)

*4.1.3 Efficiency and Ease of Use.* Participants consistently viewed efficiency and reliability as a trade-off in their information-seeking behaviors. While AI may generate wrong answers, it provides instant answers and is easily accessible. OKCs often require users to wait hours or even days for responses. Interview data reinforced this trend, revealing that users preferred utilizing AI when their priority was speed, often at the expense of depth and reliability.

A1-A3 noted that they used AI "for a quick starting point" but turned to community posts "only when the correctness really matters." A5 echoed this sentiment, stating, *"AI answers are fast, but I can't verify who said it. On Stack Overflow, at least I can read the discussion and citations below... But honestly, I am not striving for accuracy all the time; sometimes I just need to have a quick glance at a question that suddenly comes to mind."* A9, who described herself as heavily reliant on AI, succinctly captured the trade-off: *"I like AI for quick responses; you know, sometimes humans won't answer me as quickly... Reddit may even delete my post if the moderators have suspicious on my content, but there are certain moments that I really need an instant response and care."*

### 4.2 From Comparisons to Challenges (RQ1)

We compared AI and OKCs, and it is evident that AI presents mroe significant obstacles that far outweigh the decreased platform traffic. As users navigate this evolving technology, it raises critical questions about the deeper challenges and impacts underlying the declining perceptions of OKCs. Understanding these challenges not only sheds light on the current landscape but also aids in identifying effective workarounds and anticipating future implications for knowledge-sharing communities.After verifying with our interviewees, we used comics to visually represent and recapture the insights shared by them, enhancing understanding and engagement with the content.

*Obstacle 1: Convenience vs. Reciprocity (Figure 3).* The ease of using AI has led to a shift towards one-sided consumption among users, which in turn affects the culture of OKCs. As A10 noted, *"People are less inclined to engage when AI can provide quick answers."* This trend diminishes the collaborative spirit that characterizes OKCs.

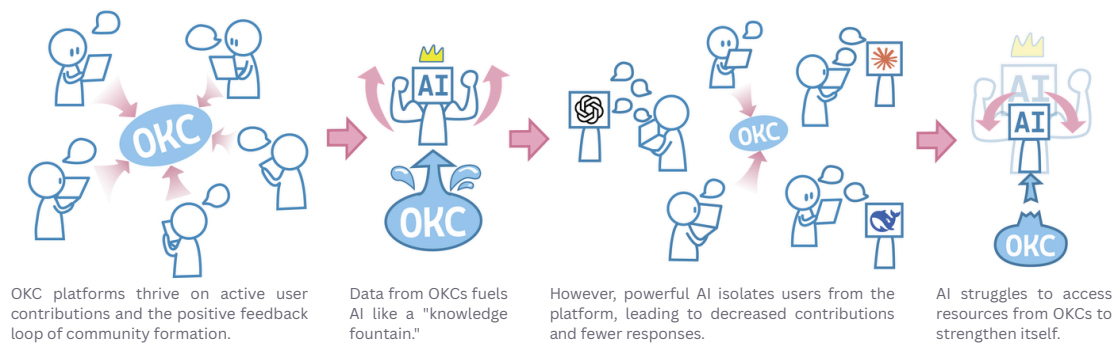

Fig. 3. The comic illustrates obstacle 1: the paradox of AI empowerment in OKCs.

*Obstacle 2: AI Hallucinations and Blurred OKCs (Figure 4).* Users often grapple with doubts about the reliability of AIGC, which is appearing more and more frequently on OKCs. A8 pointed out, *"I can't always tell if an answer is from AI or a real person, so I will never use AI,"* which blurs the boundaries and undermines confidence in OKCs. On the other hand, this uncertainty can lead users to rely more on AI, as A4 and A9 expressed: *"As I can't tell the difference, why might I not as well stick with AI for direct answers? If AI answers are trained from our online collective knowledge, what is my point to choose?"*

*Obstacle 3: Increased Workload for Moderators (Figure 5).* The convenience of AI can obscure the volunteer efforts of moderators, resulting in hidden labor and greater workloads. A10 remarked, *"Moderators are stretched thin, and it's harder to maintain trust and credibility... Frankly speaking, my friends and I can't promise the content we moderate is from human or AI users. Among us, we called this an actual Rashomon* [11]*."* This growing burden can impact the quality of oversight in OKCs.

*Obstacle 4: Platform Partnerships with AI (Figure 6).* When platforms choose to collaborate with AI, it can diminish the motivation of volunteers, as they perceive that the platform benefits financially from their contributions. A11, who considered himself the biggest dissenter of AI, noted, *"It feels like the platform profits while we do the hard work... I create*

---

[11] "Rashomon" refers to the Rashomon effect, a term for when different people provide contradictory, yet plausible, accounts of the same event.

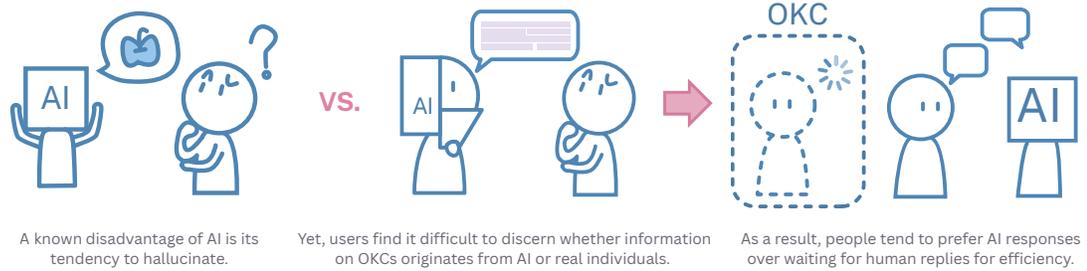

Fig. 4. The comic illustrates obstacle 2: the challenge of distinguishing AI from human contribution.

*content, and they sold mine. I feel being betrayed and am like the hidden labor without salary,"* raising concerns about privacy and the ethical implications of such partnerships.

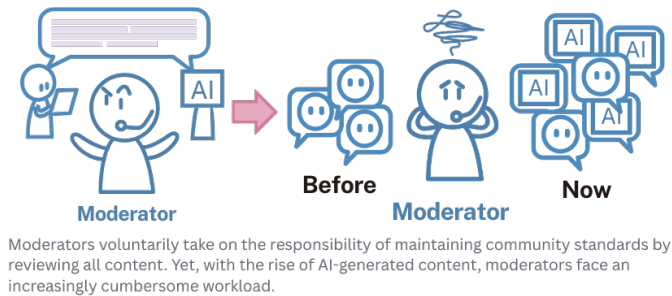
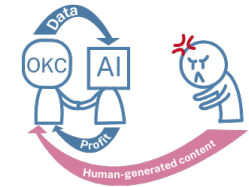

Fig. 5. The comic illustrates obstacle 3: the increasing burden on moderators.

Fig. 6. Obstacle 4: unpaid contribution.

### 4.3 Acknowledging and Working Around Limitations of OKCs (RQ1 and RQ3)

*4.3.1 Current Workarounds by Platforms.* Platforms are not passive amid the rise of AI. They are adjusting product mechanics, governance, and monetization to retain participation and maintain content quality. Our reviews, aligned with interviewees' descriptions, highlighted four key approaches (P1-P4), captured through experiences and examples from OKC. Overall, platforms demonstrated a positive attitude and are inclined to cooperate.

*P1: Increasing User Engagement.* Platforms are tightening feedback loops and lowering participation barriers. Interviewees mentioned nudges, streaks, and prompts matched to interests. A2 noted, *"I get weekly reminders with threads I once viewed but never answered—it sometimes pushes me to reply."* Features such as one-click reactions and "unanswered near you" tasks encourage lightweight involvement. Quora, Reddit, and Medium now use follow-up pings and "update old answer" prompts to sustain small but frequent engagement.

A complementary tactic is promoting proprietary apps and direct notifications that bypass search engines and AI summaries. A3 said, *"On the app, I go straight to my tags and drafts ... I skip search results now filled with AI summaries."* Push alerts and draft continuity have boosted retention, similar to how news or media apps maintain daily user via emails. These efforts establish habitual, personalized return loops critical for sustaining attention.

*P2: Diversifying Content Formats.* Communities are broadening how knowledge is presented to convey expertise that static text cannot, including adding short videos, annotated screenshots, and interactive examples. A4 observed, *"AI can discuss topics... but step-by-step photos or videos make a difference."* A5 added, *"Short clips help me grasp answers faster; I see more of them now than before."*

Since multimedia can obscure sourcing, A1 cautioned, *"Without transcripts and citations, videos feel less verifiable than text."* A6 also noted such trend on Reddit's videos where he noted, *"I remember there are some videos with translation now."* To preserve credibility, some communities pair clips with reproducible examples, editable code blocks, and linked sources. CNN and Wikipedia demonstrate similar adaptations—redirecting consumers to controlled environments for full, cited access. These format experiments showcase how OKCs may compete through transparency and tacit, contextual knowledge.

*P3: Expanding Revenue Channels.* As ad income and search referrals decline, OKCs are turning to subscriptions and enterprise products. A4 explained, *"The pro tier bundles ad-free reading and advanced search… I pay, but I worry good answers may hide behind paywalls."* A2 agreed that fees are acceptable *"if they fund moderation and upkeep—just not visibility bias."* When asked about monetization, A10, a Reddit moderator, joked, *"if the platform makes more money and we get a little cut, who would reject doing more moderation?,"* reflecting latent hope for fair recognition of community labor. Stack Overflow's "Teams" product exemplifies mixed models that fund expert review while keeping archives open. Across accounts, respondents framed acceptable monetization as reinvestment in trust and quality rather than to charge high fees and sharply restrict access.

*P4: Partnering or Licensing with AI Platforms.* Some platforms are exploring formal agreements with AI companies. A3 remarked, *"I noticed an 'AI-assisted' badge and links to sources... (it) helps to see where text came from."* Respondents favored explicit attribution, backlinks, and transparency on AI use. A1 added, *"If an AI summary sits atop a thread, it should cite specific answers; otherwise it feels parasitic."*

While these are not typical OKC platforms, examples such as Wikipedia's licensing agreements with search tools and Bloomberg's data partnerships indicate possible business models through regular corpus licensing, link requirements, or paid attribution. In addition to licensing, some participants believed that datasets or analytic reports are more ideal than selling all discussions as revenue streams, because these alternatives preserve privacy and visibility.

*4.3.2 Current Workarounds by Users.* Three major themes (U1-U3) are concluded.

*U1: Calls for Moderator Action.* Many users sought stronger moderation to curb AI misuse and maintain trust. 38 respondents explicitly wanted stricter controls in our survey's open ended questions, often phrased as "the stricter to AI the better." A2 also argued, *"Flagging AIGC should be mandatory... if a post comes from a bot, label it and require sources."* A5 valued recency and verification: *"I trust posts with 'last updated' notes; if AI answers lack that, I skip them."* On the other hand, A10 emphasized that the moderation work is becoming increasingly deminding, thus requiring a balance, although he also acknowledged that moderation is a core "what we can do" in OKCs.

Moderation also preserves social cohesion. A5 said, *"Moderators who defuse flame wars and nudge edits keep me here."* A2 added, *"When fights break out over who used AI, I just click away."* A8 recalled reading post on Stack Overflow that a user complain his answer from AI should not be deleted, saying, "thanks for deleting the post as I feel like people nowadays believe they own the work genereated from AI, even AI is actually trained by many (other's) referenced data." Overall, most users accepted AI contributions only within accountable, transparent, and civility-focused frameworks.

*U2: Reimagining Incentives and Rewards.* Users still value badges and recognition but want them aligned with quality and maintenance, not speed. A3 reflected, *"My gold badge came from curating one canonical answer over months, not quick hits."* Even small contributions mattered like A5 said, *"I'm proud of my few badges, though I only comment."*

Several proposed rewarding verification and updates rather than streaks. A5 suggested, *"Give badges for thorough sourcing or for improving old posts."* A4 added, *"If I add references and raise an answer's trust score, that should count more."* Recognition of slow, substantive labor can preserve content depth despite automation pressures.

*U3: Selective Engagement and Pragmatic AI Use.* Some participants accepted that AI will dominate general queries and adjusted their behavior accordingly. A1 said, *"If AI replaces OKC, I'll follow—it means it's capable."* A9 echoed with history, "If reliance is inevitable, I'll accept it... Don't you think the advent of AI is just as how internet, social media, and all (kinds of new technologies) appear? The public, sooner or later will accept AI as how we accept the internet." Yet many found niches that remain human-centered. A4 muted tags flooded with AI spam: *"I'd rather not engage than police threads."* A3 migrated to smaller specialist groups: "They still feel human." A8 and A11 opposed AI as "they are making up stories... not creating".

Such selective engagement suggests OKCs may endure as focused, high-trust spaces while AI satisfies generic tasks. Others combined both tools: A1 began with AI for a "sketch," then verified with community threads; A2 reversed the flow for safety-critical topics. Users thus blend immediacy and reliability across sources. As A1 concluded, *"I trust OKC because of human experience—but if that dries up, I won't pretend otherwise."* Their pragmatism highlights the need to keep communities credible and welcoming.

*4.3.3 Workarounds Imagined or Developed by Respondents.* Some participants (mainly A1, A5, and A10) envisioned a cooperative model rather than competition between AI and OKCs. A recurring idea was an AI agent acting as a bridge: when users got a helpful AI response, they could consent to sharing it. In another words, posting both the question and answer back to the community. If a related thread existed, the AI would append the answer with an "AI-generated" label; otherwise, it would start a new post. This flow (colored purple in Figure 1) would recover lost collective knowledge and keep archives alive.

A10 saw potential for mutual benefit as he noted in the moderation works, saying, *"Let's face it, AI is not bad... AI replies quickly, and I am amazed by some generated images in my subcommunity... (the one he moderates) is quite formal and scientific... not attractive to general users... I once saw generated figures and comics to illustrate the concepts and struggled to decide how to handle it, as our rule is no AIGC!"* Besides, AI offers immediacy, while the community gains curated data and visibility. Yet they warned of privacy and authorship challenges. Posting full transcripts might expose sensitive queries, and unclear crediting could undermine contributor identity. P3 and P5 stressed opt-in consent, transparent provenance, and clear moderation of AI posts to prevent erosion of community reciprocity.

### 4.4 Knowledge Seekers' Perceptions (RQ2 and RQ3)

*4.4.1 Perceptions and Practices in General.*

*AI as a Tool for Knowledge Seeking.* Survey data suggest that participants generally perceive AI as a useful but double-edged tool that is good for content creation and summarization but bad for originality, potentially because of no sourcing. As shown in Figure 7, most respondents tend to agreed that AI makes content creation easier and trust its summarization and answer-generation abilities. However, the majority disagreed with the claim that AI had "no impact on originality," showing awareness of trade-offs between convenience and creative value. In sum, respondents tend to position AI as a time-saving assistant that, nonetheless, reshapes the meaning of originality, authorship, and contribution within the knowledge exchange system.

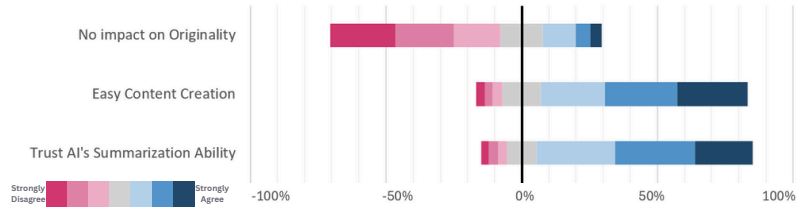

Fig. 7. Respondents' perception on AI as knowledge seeking tools.

*AI as Perceived Complementarities and Tensions in OKC.* To better capture this ambivalence, survey items were grouped into three thematic clusters.

The first cluster shows broad consensus on AI's potential negative impacts. Participants tended to agree that AIGC often lacks originality (Q31), may contain bias or ethical risks (Q32), and could discourage user participation (Q36，38–Q39). More than half of respondents preferred to avoid AI-generated material in posts (Q28) and worried that its proliferation diminishes users' motivation to engage. The stacked distributions in Figure 8 illustrate a clear concentration toward the "agree" end, colored in blue, underscoring community apprehension that GenAI could erode authenticity and human exchange.

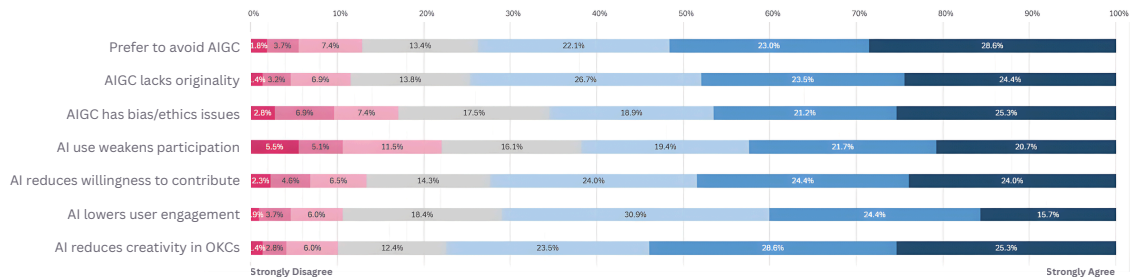

Fig. 8. Respondents' concerns about AI's negative impact.

Our second cluster reveals nuanced perceptions surrounding AI literacy and responsibility (Figure 9). While many respondents claimed competence in identifying AI text (Q26), responses were largely neutral to modestly positive about their ability to consistently do so, indicating uncertainty about detection reliability. Around half admitted to ignoring suspected AIGC (Q27), reflecting practical fatigue rather than categorical rejection. This suggests AIGC may increase the cognitive load among OKCs users. Meanwhile, high agreement levels (Q41, Q42) emphasize the importance of maintaining OKCs and the risk that their decline could harm AI's own development, suggesting a mutual dependence recognized by respondents. Respondents seemed aware that AI's progress ultimately relies on the quality and sustainability of the human knowledge base it draws from, further confirmed in the discussion with our interviewees.

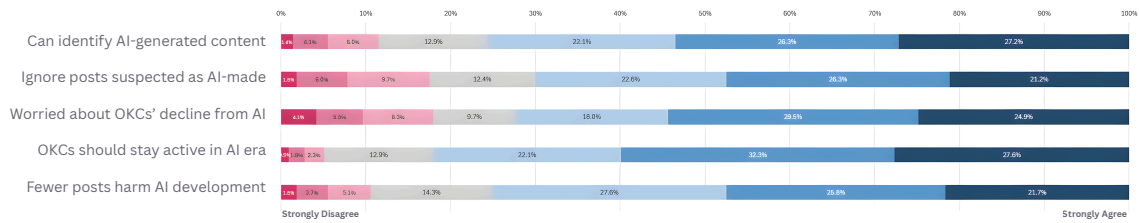

Fig. 9. Responses related to AI literacy and responsibility.

Our last cluster addresses trust and the perceived positive influence of AI. Although some agreed that AI can contribute useful content (Q29, Q30, Q34), aligning with the results on AI as a time-saving assistant previously, nearly half rated AI quality below human-created standards (Q35). The majority remained neutral or cautiously optimistic rather than enthusiastic, signaling measured acceptance. Figure 10 reflects this ambivalence with a predominance of light blue and gray bands, as participants negotiated between pragmatic use of AI outputs and skepticism toward their depth and reliability.

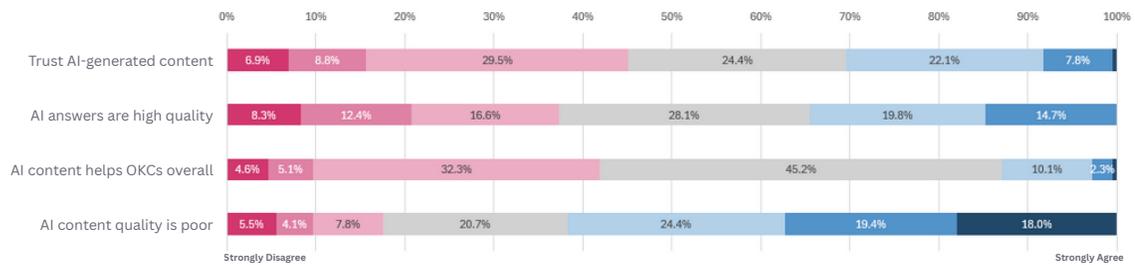

Fig. 10. Respondents' perception of trust and the positive impact of AI on OKCs.

In short, survey responses showed a balanced yet ambivalent pattern, with few strong consensuses. Participants generally accept AI for its functional value in information retrieval and summarization but remain protective of the social and epistemic foundations that sustain OKCs. Most favor cautious coexistence over replacement. While the majority of respondents reported that they can identify in between human The following interview insights further illustrate how such diverse perceptions may stem from participants' different roles within OKCs.

*4.4.2 Moderators.* The single moderator interviewed (A10) adopted a somewhat pragmatic and neutral tone, focusing less on ideological debates and more on practical governance implications. He acknowledged both the convenience and complications AI brings to community management, remarking that AI can quickly surface duplicate or low-quality content but also floods moderators with questionable, synthetic posts. His view highlights awareness of unpaid labor and moderation burnout—issues intensified by AI-generated noise.

Yet A10's stance cannot represent moderators universally. In fact, the other moderator replied in our survey and later rejected the interview was negative towards the use of AI, writing in the email, *"Reddit is my pure and unpolluted land and AI is contaminating it."* In this study, A10 stands as an indicator rather than a full sample: moderation perspectives remain underrepresented but are likely critical to understanding how OKCs will handle AI governance at scale.

*4.4.3 Content Creators (Questioners and Commenters)．* Content creators can be considered as deeper engagement in the community who pose and answer questions (e.g., A3, A5, A7, A11). They were generally more skeptical and critical toward AI. Their skepticism stemmed less from fear of technological takeover than from a perceived threat to credibility, identity, and the social capital they had built within the community.

Experienced contributors (A3, A5) worried that AI-generated responses could "flatten" expertise and diminish recognition for high-quality human contributions. A7, who mainly comments, noted that AI replies often lack context and may discourage further discussion, while A11 directly rejected the use of AI, arguing that it undermines authenticity and trust.

Despite these concerns, some (e.g., A3, A5) acknowledged AI's utility as a preliminary aid. They regarded AI as a helpful tool for quick overviews but inadequate without human interpretation and validation. Overall, content creators displayed a more guarded pragmatism. They draw efficiency from AI but actively defend their epistemic legitimacy and interpersonal bonds within OKCs.

*4.4.4 Content Users/Viewers.* In contrast, less engaged users (viewers who primarily browse or consume content) expressed more favorable and adaptive attitudes toward AI integration. This group, including A1, A2, A4, A6, A8, and A9, generally perceived AI as an enabling tool that enhances accessibility and efficiency rather than a disruptive force. Their views align with the survey trends showing neutral-to-positive responses toward AI quality and usefulness, particularly in relation to convenience and knowledge retrieval.

Several participants appreciated AI's ability to condense long discussions and present key insights without extensive searching. A4 tended to agree with integrating AI, viewing it as a natural continuation of technological progress. A9 expressed even stronger enthusiasm, likening AI's emergence to the rise of the early internet in the 2000s as an inevitable transformation that democratizes knowledge and broadens participation. She compared the experience and emphasized that instant response is a gem that human users cannot achieve. A6 and A8 similarly highlighted AI-generated summaries as effective gateways for quick learning, especially for users with limited time or domain expertise.

These participants valued speed, accessibility, and simplicity over authorship or prestige, often framing GenAI as a complement that augments, rather than replaces, human contributions. As A2 observed, *"AI helps me find relevant threads faster; I still read the top answers if I want details."* While most users in this group acknowledged potential risks such as misinformation and bias, they expected platforms or moderators (rather than individuals) to ensure content reliability.

## 5 Discussion

Our study draws inspiration from Preece's seminal work in the HCI community [60], which questions how online communities can be systematically evaluated for both usability and sociability. With the emergence of AI significantly influencing the sociability of our communities, we investigate the newly formed human-AI interactions within OKCs. We now reflect on our findings and examine how these interactions, alongside socio-technical design, impact sustained participation, particularly in light of recent advancements in AI.

### 5.1 Comparing User Experience with AI and OKCs (RQ1)

AI and OKCs now coexist as parallel infrastructures for knowledge seeking, each offering distinct affordances. Participants described AI as a space for immediacy and accessibility, and OKCs as environments for contextual quality and peer validation. This functional differentiation mirrors recent large-scale evidence showing reduced engagement in developer and question-answering communities following AI's public adoption [10]. The tension between rapid efficiency and communal value signals not a competition of platforms but a reallocation of users' cognitive and temporal investments. While AI addresses utilitarian queries with unmatched speed, OKCs should sustain interpretive depth and sociability.

Our quantitative results demonstrated significant reductions in answering, commenting, and question-asking activities, aligning with qualitative accounts that portrayed ChatGPT and similar tools as the "first stop" for uncertainty. Yet rather than a wholesale abandonment, many users described hybrid workflows that layered both — using AI for exploration and OKCs for verification and expertise. This layering indicates a subtle reconfiguration of epistemic trust. AI may outperform OKCs in accessibility, but communities continue to function as an anchor for credibility. Such triangulation reinforces that OKCs remain critical where social cues, plurality of viewpoints, and tacit knowledge matter. These findings extend prior work on participation motivation [25, 31] by showing how convenience reframes reciprocity. As personal search becomes frictionless, the social value of contributing back diminishes. The erosion of "give-and-take" norms observed here resonates with broader concerns in online collaboration studies that automation can alter not only volume but also the moral infrastructure of contribution [33]. Maintaining OKCs will therefore depend on cultivating renewed incentives that compete not in speed but in meaning, moderation, recognition, and peer connection; aligning the sociality mentioned by Preece [60].

### 5.2 Questioning Authenticity Under Automation (RQ1)

Questions of authenticity emerged as the most salient cross-cutting concern. Users simultaneously appreciated AI's generative capacity and doubted its veracity, unveiling an epistemic paradox: AI democratizes access while destabilizing provenance. Even most of our survey respondents believed they were confident enough to judge the difference between AIGC and human-written content, but interviewees often struggled to discern whether posts were written by humans or machines, echoing findings that AIGC can evade accurate human detection [57]. This blurring challenges the relational trust that underpins OKCs. As seen in our interviews, when contributors suspected automation, they either disengaged or requested stricter moderation—signals of declining confidence in communal integrity.

The rise of AI-generated answers within OKCs also complicates authorship and ownership. Some framed AIGC as contamination, while others viewed it as an opportunity for augmentation. These divergent stances align with recent analyses of subreddit rule evolution showing that AI-related rules have rapidly proliferated, particularly in large, media-oriented communities, as moderators attempt to codify authenticity at scale [42]. Complementary qualitative work has documented similar enforcement tensions: even where communities restrict AI content, moderators rely on time-intensive heuristics and context clues to judge compliance [43]. Our study thus situates user and moderator skepticism within a broader pattern of emergent, imperfect governance.

Authenticity also constitutes an affective boundary between algorithmic and human knowledge. Participants valued the experiential richness and dialogue inherent to OKC posts—qualities that generative models mimic syntactically but not socially. The perception that AI "sounds certain but lacks lived context" underscores the limitation of synthetic fluency as a substitute for accountability. As research in AI-mediated communication notes [28, 31], the suspicion of automation can itself diminish interpersonal trust, regardless of actual AI involvement. We question whether this suspicion undermines OKCs, which are fundamentally motivated by learning through shared experiences and emotional reciprocity [59, 77, 78].

### 5.3 The Thousand "Hamlets" in Online Knowledge Communities (RQ2)

Each participant, creator, moderator, or even the platform confronts a distinct mix of opportunity and disquiet, thus forming an uninformed, heterogeneous moral landscape.

For knowledge seekers, especially among lower engagement group at OKCs, AI constitutes a pragmatic assistant. They prize its convenience and summarize it as "good enough," foreshadowing an echo of utilitarian trust. These users adapt effortlessly, treating OKCs as secondary verification channels. In contrast, established content creators perceive existential stakes. They fear dilution of expertise and the devaluation of human contribution, since AI-generated posts often replicate surface structures of credibility without reputational labor. This sentiment aligns with prior work on expertise erosion in crowdsourced systems [45]. For moderators, AI introduces procedural ambiguity like "*Rashomon*." Identifying, labeling, and deciding on AI posts multiplies emotional and cognitive load.

These coexisting "Hamlets" reveal plural, role-dependent interpretations of authenticity and progress. The coexistence of technophilia (efficiency, novelty) and technophobia (contamination, labor exploitation) reflects a broader socio-technical negotiation. Importantly, the appearance of polarization does not imply disorder but adaptation. OKCs are recalibrating normative boundaries and deciding what counts as expertise, authorship, and fair use at unprecedented speed and scale. Such pluralism, if channeled productively, can inform more participatory rule-making that centers community self-determination rather than centralized enforcement. That said, the current centralized workarounds from platforms are not advocated.

### 5.4 Developing Workarounds with AI (or Not)(RQ2)

Despite significant strain, both platforms and users are already experimenting with workarounds to stabilize the AI–OKC relationship. Our reviews show that communities are not passive victims of technological disruption but active designers of response mechanisms.

At the platform level, interventions include nudging engagement, diversifying media forms, introducing subscription models, and formalizing partnerships with AI companies. These strategies reflect efforts to rebuild sustainable attention loops and financial independence as referral traffic and ad revenue decline. Elements such as "answer reminders," "update prompts," and content-format diversification directly respond to users' preference for immediacy while preserving communal discourse. However, these design tactics risk reorienting participation around metrics rather than meaning. The challenge lies in ensuring that automation augments moderation and curation, rather than reducing interaction to gamified repetition.

Partnering with AI vendors offers both opportunity and peril. Transparent data licensing and attribution could recognize community labor and generate fair compensation, but opaque deals risk deepening disillusionment among contributors who feel their unpaid work is exploited for commercial AI training [12]. Anxieties over data ethics and ownership arise. HCI research can contribute by developing governance models that encode consent, visibility, and fair redistribution into data-sharing pipelines.

From the user side, grassroots adaptations are equally revealing. Many participants advocated mandatory labeling of AI-generated text and rewarded moderation transparency. Others called for new incentive structures—badges for verification, maintaining rather than producing content—signaling a shift from productivity to stewardship values. Participants also imagined cooperative systems where AI acts as a bridge, posting its own generated solutions back to the community for review and archival. Such hybrid arrangements could reclaim dispersed knowledge while giving communities oversight. Yet, without robust consent and attribution protocols, they risk perpetuating the same asymmetries that now trouble contributors.

These observations underscore that sustaining OKCs under automation requires more than technical fixes. It demands rearticulation over the social contract of online knowledge exchange, including but not limited to acknowledging invisible labor, clarifying data provenance, and rewarding participation that upholds credibility rather than volume. We suspect platforms that center co-governance, where moderators and users define AI policies collectively, are better positioned to balance efficiency with authenticity. The rapid proliferation of AI-specific rules on Reddit exemplifies this emerging self-moderation capacity [42], but operational support remains limited. Future research should explore participatory design frameworks enabling communities to prototype their own AI-use charters, supported by detection tools and ethical templates rather than top-down mandates.

### 5.5 Design Implications (RQ3)

Our study highlights how the introduction of GenAI has complicated the ecology of OKCs. Participants described obstacles distributed across platforms and users, but also surfaced tentative or speculative workarounds. Drawing on these themes, we discuss several implications for the design of future OKCs and their intersections with AI.

Our findings reveal significant deficiencies in the transparency of AIGC on the current OKCs platform. Although participants acknowledged that iterative advancements in technology have improved the quality of AIGC, the platform still exhibits clear shortcomings in terms of transparency and disclosure of content provenance. Participants consistently reported that, due to concerns regarding content credibility, the authenticity of social connections, and the risk of potential manipulation, they actively or passively engage in judgments about whether unmarked content is human- or AI-authored. This additional cognitive load exerts a negative impact on community engagement and trust. We argue that platforms should enforce greater transparency and disclosure of AIGC—through measures such as mandatory labeling, monitoring and disclosure requirements for accounts that frequently disseminate AIGC, or similar mechanisms to relieve users of this cognitive burden.

In addition, our study reveals that community members exhibit a strong dependence on volunteer moderators, expecting them to assume primary responsibility for verifying the provenance and quality of content. However, this critical labor is entirely uncompensated. Volunteer moderators themselves report that the proliferation of AIGC has substantially increased their workload, to the extent that focusing on policing the flood of AIGC now contravenes the original ethos of community support and mutual aid. The issues of uncompensated passive labor and inequitable distribution of responsibility have become far more pronounced and untenable in the AI era. We therefore contend that platforms must proactively assume the primary responsibility for content governance rather than outsourcing it to volunteers. At the same time, platforms should actively respect, value, and support the contributions of community moderators by collaborating with them in the co-design of content evaluation mechanisms and community guidelines, thereby establishing a more equitable and sustainable moderation ecosystem.

Beyond complete reliance on moderators, our findings also reveal opportunities to reconfigure incentive mechanisms to sustain human contributions. Although creators and reviewers feel threatened by displacement from artificial intelligence tools, they still highly value status and reputation when their efforts are recognized. Existing research stresses that sustaining communities requires strengthening prosocial motivations such as reciprocity and acknowledgment[11, 39]. Specifically, designers could extend reputation systems to distinguish certified human-authored answers from AI-assisted content, thereby ensuring the visibility and prestige of authentic contributions. Similarly, platforms could adopt "community-first" reward mechanisms that recognize human review of AI-submitted content as a form of knowledge curation.

Our findings reveal that stakeholders of different identities perceive AIGC on the OKCs platform in divergent ways. Although certain predictable trends exist, the challenge of balancing diverse user needs remains an issue that cannot be overlooked. Currently, this issue is addressed entirely through varying rules established by different sub-communities or through users' alternative uses of existing features. As noted earlier, the enforcement of these rules relies heavily on community members' self-discipline and the voluntary labor of moderators. In a context where AI generation costs are extremely low and AIGC is significantly more controversial than human-created content, the operation of these rules is highly unstable. We argue that the platform should proactively respond, at the level of feature design, to the common needs reflected in sub-community rules, such as one-click blocking of AIGC and prioritizing the display of user-contributed content.

Finally, we emphasize the importance of considering **long-term community resilience**. Although some users express helplessness about OKCs potentially declining due to AI replacement, most participants accept using AI as a retrieval tool, especially for its advantages in timeliness and search breadth, while still affirming the superior credibility and quality of human replies on OKCs. We therefore argue that future platform design should fully account for the respective ecological niches of AI and real users within the community, achieving human–AI collaboration rather than simple substitution through systematic resilience mechanisms. For time-sensitive technical questions, AI could first rapidly generate candidate replies, which would then be reviewed, certified, and archived into the knowledge base by community users. This workflow echoes design proposals that position AI as a collaborator rather than a replacer [68], highlighting human agency in deciding what information enters the community record. It leverages AI's efficiency advantages while maintaining the community's collective control over trustworthy knowledge—under the premise of preserving autonomy through data governance practices developed in cooperation with AI companies, as well as governance structures that enable the community to dynamically adapt to the balance between human activity and machine assistance.

In sum, design for hybrid knowledge ecologies must go beyond accommodating AI technologies at the surface level. Instead, effective interventions will need to foreground transparency, equitable burden-sharing in platform governance, hybrid workflows, and resilience. We dedicate these implications chart possible pathways for sustaining OKCs not in opposition to GenAI, but in productive tension with it.

## 6 Miscellaneous

### 6.1 Acknowledgments of the Use of AI

We utilized the Large Language Model (LLM), specifically ChatGPT 4.o, in the preparation of this manuscript. Their role was strictly limited to assisting with word choices during translation, formatting issues, and grammar correction. The core research ideas, experimental design, data analysis, findings, discussions, and conclusions presented in this paper are entirely authors' original work and were not generated by AI. All AI-assisted content was carefully reviewed, verified, and edited by the authors to ensure accuracy. The authors take full responsibility for the final content of this paper.

22 Pang et al.## 6.2 Limitations and Future Work

Our mixed-method study combined interviews, surveys, reviews, and content analysis to examine how AI influences participation, authenticity, and governance within OKCs. While this multi-method approach allowed us to capture perspectives from diverse community roles, it also introduces limitations that affect the generalizability and temporal relevance of our findings.

First, we focused on communities that had already begun discussing or implementing AI-related rules. This sampling decision likely overrepresents communities where AI is a visible or contested topic and underrepresents those without formal AI policies. For instance, several subcommunities from Reddit (e.g., *r/appropriatetechnology*, *r/Human_Artists_Info*, *r/ArtistLounge*, and *r/ArtistHate*) reflect active debates about AI's influence on creative work. Our recruitment primarily drew from medium to large communities (average membership over 50,000), which tend to have established moderation teams and clearer rule structures. As a result, the findings may differ from dynamics in smaller or emerging communities that operate with fewer resources and closer interpersonal networks. Future research should specifically examine these smaller groups to explore how resource constraints and community intimacy affect AI adoption, rule enforcement, and perceptions of authenticity.

Second, all interviews were conducted between mid- and late-2025, a period marked by rapid technological advancement in GenAI and shifting public awareness. Our limited sample size ($N = 11$) may not fully capture the evolving diversity of user perceptions and behaviors. Moreover, participant roles were unevenly distributed, with only one moderator represented, which may bias our interpretation of governance dynamics. Although our qualitative coding achieved strong inter-coder reliability, interpretive biases inherent in retrospective accounts remain possible. To enhance validity, future research could include longitudinal designs or direct ethnographic observation of moderation processes involving AI-generated posts. Additionally, complementing qualitative accounts with server-side behavioral logs or network analyses could help link participants' self-reported experiences to actual engagement patterns and content flows within OKCs.

## 7 Conclusion

This paper examines how GenAI is reshaping OKCs by combining a survey of 217 participants with interviews of 11 current OKCs users. OKC, as vital components of HCI and CSCW research, play a crucial role in enabling collaborative knowledge exchange. The impact of AI is a pressing and inevitable issue for these communities. Participation is shifting from contribution to consumption as AI's convenience weakens long-standing reciprocal norms. Authenticity, once grounded in transparent peer dialogue, is now strained by indistinguishable machine-generated content, placing new burdens on users and volunteer moderators. Respondents express polarized views toward AI, yet regardless of stance, sustaining sociability, empathy, and reciprocity remains essential for building resilient, human-centered knowledge ecologies. By framing decline not as inevitable replacement but as a design challenge of sustaining socio-technical complementarities, we highlight opportunities for HCI to imagine resilient infrastructures that balance AI efficiency with the communal value of human judgment. ***To decline or sustain, it largely depends on how we act. We call for more innovative procedures to sustain our communities***.